\pdfoutput=1
\documentclass[runningheads]{llncs}

\usepackage[utf8]{inputenc}
\usepackage[T1]{fontenc}

\usepackage{ifthen}
\ifthenelse{\isundefined{\drafttrue}}{
  \usepackage[disable]{todonotes}
}{
  \usepackage{showlabels}
  \usepackage{todonotes}
}

\usepackage{amsfonts}
\usepackage{amsmath}
\usepackage{amssymb}

\usepackage{tikz}
\usetikzlibrary{automata,positioning,fit,calc}

\usepackage{algorithm2e}[ruled]

\usepackage{graphicx}
\usepackage{color}

\usepackage{float}

\usepackage{listings}
\usetikzlibrary{tikzmark}

\let\oldforall\forall
\renewcommand{\forall}{\oldforall \,}
\let\oldexists\exists
\renewcommand{\exists}{\oldexists \,}

\newcommand{\integers}{\mathbb{N}}
\newcommand{\isep}{\mathrel{{.}\,{.}}\nobreak}
\newcommand{\bigo}{\mathcal{O}}

\newcommand{\np}{\textsf{NP}}
\newcommand{\ap}{\mathsf{AP}}
\newcommand{\kripke}{\mathcal{K}}
\newcommand{\runs}{\mathcal{R}}

\newcommand{\rdiameter}{\alpha}

\newcommand{\contraction}{\mathcal{G}}

\newcommand{\charnum}{\mathcal{C}}
\newcommand{\ltl}{\textsf{LTL}}
\newcommand{\ltlf}{\textsf{LTL}_\textsf{f}}
\newcommand{\ctl}{\textsf{CTL}}
\newcommand{\ctla}{{\textsf{CTL}_{\oldforall}}}
\newcommand{\ctlu}{{\textsf{CTL}_{\until}}}
\DeclareMathOperator{\genop}{{\triangleright}}
\DeclareMathOperator{\ax}{{\oldforall\mathsf{X}}}
\DeclareMathOperator{\ex}{{\oldexists\mathsf{X}}}
\DeclareMathOperator{\nxt}{\mathsf{X}}
\DeclareMathOperator{\af}{{\oldforall\mathsf{F}}}
\DeclareMathOperator{\ef}{{\oldexists\mathsf{F}}}
\DeclareMathOperator{\finally}{\mathsf{F}}
\DeclareMathOperator{\ag}{{\oldforall\mathsf{G}}}
\DeclareMathOperator{\eg}{{\oldexists\mathsf{G}}}
\DeclareMathOperator{\globally}{\mathsf{G}}
\DeclareMathOperator{\until}{\mathsf{U}}
\DeclareMathOperator{\au}{{\oldforall\mathsf{U}}}
\DeclareMathOperator{\eu}{{\oldexists\mathsf{U}}}

\newcommand{\tree}{\mathcal{T}}
\newcommand{\sdag}{\mathcal{D}}
\newcommand{\labels}{\mathcal{L}}

\newcommand{\leftv}{\mathsf{l}}
\newcommand{\rightv}{\mathsf{r}}
\newcommand{\psample}{S^+}
\newcommand{\nsample}{S^-}
\newcommand{\sample}{\psample, \nsample}
\newcommand{\learn}{\textsf{L}}
\newcommand{\minlearn}{\textsf{ML}}
\newcommand{\ssample}{S}
\newcommand{\spsample}{S^+}
\newcommand{\snsample}{S^-}

\newcommand{\sepfor}{\mathcal{S}}
\newcommand{\sat}{\textsf{SAT}}
\newcommand{\rank}{\rho}

\title{SAT-based Learning of Computation Tree Logic}

\author{Adrien Pommellet\inst{1}\orcidID{0000-0002-1825-0097} \and
  Daniel Stan\inst{2}\orcidID{0000-0002-4723-5742} \and
  Simon Scatton\inst{3}}

\institute{LRE, EPITA, France \\
  \email{adrien@lrde.epita.fr} \\
  \url{https://www.lrde.epita.fr/~adrien/} \and
  LRE, EPITA, France \\
  \email{daniel.stan@epita.fr} \\
  \url{https://www.tudo.re/daniel.stan/} \and
  LRE, EPITA, France \\
  \email{simon.scatton@lrde.epita.fr}}

\begin{document}

  \maketitle

  \begin{abstract}
    The {\ctl} learning problem consists in finding for a given sample of positive and negative Kripke structures a distinguishing {\ctl} formula that is verified by the former but not by the latter. Further constraints may bound the size and shape of the desired formula or even ask for its minimality in terms of syntactic size. This synthesis problem is motivated by explanation generation for dissimilar models, e.g. comparing a faulty implementation with the original protocol. We devise a {\sat}-based encoding for a fixed size {\ctl} formula, then provide an incremental approach that guarantees minimality. We further report on a prototype implementation whose contribution is twofold: first, it allows us to assess the efficiency of various output fragments and optimizations. Secondly, we can experimentally evaluate this tool by randomly mutating Kripke structures or syntactically introducing errors in higher-level models, then learning {\ctl} distinguishing formulas.
    
  \end{abstract}

  \section{Introduction}
  
  \emph{Passive learning} is the act of computing a theoretical model of a system from a given set of data, without being able to acquire further information by actively querying said system. The input data may have been gathered through monitoring, collecting executions and outputs of systems. Automata and logic formulas tend to be the most common models, as they allow one to better explain systems of complex or even entirely opaque design.
  
  \emph{Linear-time Temporal Logic} {\ltl}~\cite{Pnueli77} remains one of the most widely used formalisms for specifying temporal properties of reactive systems. It applies to finite or infinite execution traces, and for that reason fits the passive learning framework very well: a {\ltl} formula is a concise way to distinguish between correct and incorrect executions. The {\ltl} learning problem, however, is anything but trivial: even simple fragments on finite traces are \np-complete~\cite{Fijalkow21}, and consequently recent algorithms tend to leverage {\sat} solvers~\cite{Neider18}.
  
  \emph{Computation Tree Logic} {\ctl}~\cite{Clarke81} is another relevant formalism that applies to execution trees instead of isolated linear traces. It is well-known~\cite[Thm. 6.21]{PoMC} that {\ltl} and {\ctl} are incomparable: the former is solely defined on the resulting runs of a system, whereas the latter depends on its branching structure.
  
  However, the {\ctl} passive learning problem has seldom been studied in as much detail as {\ltl}. In this article, we formalize it on \emph{Kripke structures} (KSs): finite graph-like representations of programs. Our goal is to find a {\ctl} formula (said to be \emph{separating}) that is verified by every state in a positive set $\psample$ yet rejected by every state in a negative set $\nsample$.
  
  We first prove that an explicit formula can always be computed and we bound its size, assuming the sample is devoid of contradictions. However, said formula may not be minimal. The next step is therefore to solve the bounded learning problem: finding a separating {\ctl} formula of size smaller than a given bound~$n$. We reduce it to an instance $\Phi_n$ of the Boolean satisfiability problem whose answer can be computed by a {\sat} solver; to do so, we encode {\ctl}'s bounded semantics, as the usual semantics on infinite executions trees can lead to spurious results. Finally, we use a bottom-up approach to pinpoint the minimal answer by solving a series of satisfiability problems. We show that a variety of optimizations can be applied to this iterative algorithm. These various approaches have been implemented in a \texttt{C++} tool and benchmarked on a test sample.
  
  \paragraph{Related work.}
  
  Bounded model checking harnesses the efficiency of modern {\sat} solvers to iteratively look for a witness of bounded size that would contradict a given logic formula, knowing that there exists a completeness threshold after which we can guarantee no counter-example exists. First introduced by Biere et al.~\cite{Biere99} for {\ltl} formulas, it was later applied to {\ctl} formulas~\cite{Penczek02,Xu09,Zhang09}.
  
  This approach inspired Neider et al.~\cite{Neider18}, who designed a {\sat}-based algorithm that can learn a {\ltl} formula consistent with a sample of \emph{ultimately periodic} words by computing propositional Boolean formulas that encode both the semantics of {\ltl} on the input sample and the syntax of its Directed Acyclic Graph (DAG) representation. This work spurred further {\sat}-based developments such as learning formulas in the property specification language \texttt{PSL}~\cite{Roy21} or $\ltlf$~\cite{Camacho21}, applying \texttt{MaxSAT} solving to noisy datasets~\cite{Gaglione21}, or constraining the shape of the formula to be learnt~\cite{Lutz23}. Our article extends this method to {\ctl} formulas and Kripke structures. It subsumes the original {\ltl} learning problem: one can trivially prove that it is equivalent to learning {\ctl} formulas consistent with a sample of lasso-shaped KSs that consist of a single linear sequence of states followed by a single loop.
  
  Fijalkow et al.~\cite{Fijalkow21} have studied the complexity of learning $\ltlf$ formulas of size smaller than a given bound and consistent with a sample of \emph{finite} words: it is already $\np$-complete for fragments as simple as $\ltlf(\land, \nxt)$, $\ltlf(\land, \finally)$, or $\ltlf(\finally, \nxt, \land, \lor)$. However, their proofs cannot be directly extended to samples of infinite but ultimately periodic words.
  
  Browne et al.~\cite{Browne88} proved that KSs could be characterized by {\ctl} formulas and that conversely bisimilar KSs verified the same set of {\ctl} formulas. As we will show in Section \ref{sec:learning}, this result guarantees that a solution to the {\ctl} learning problem actually exists if the input sample is consistent.
  
  Wasylkowski et al.~\cite{Wasylkowski11} mined {\ctl} specifications in order to explain preconditions of \texttt{Java} functions beyond pure state reachability. However, their learning algorithm consists in enumerating {\ctl} templates of the form $\af a$, $\ef a$, $\ag (a \implies \ax \af b)$ and $\ag (a \implies \ex \ef b)$ where $a,b \in \ap$ for each function, using model checking to select one that is verified by the Kripke structure representing the aforementioned function.
  
  Two very recent articles, yet to be published, have addressed the {\ctl} learning problem as well. Bordais et al.~\cite{Bordais2023} proved that the passive learning problem for {\ltl} formulas on ultimately periodic words is {\np}-hard, assuming the size of the alphabet is given as an input; they then extend this result to {\ctl} passive learning, using a straightforward reduction of ultimately periodic words to lasso-shaped Kripke structures. Roy et al.~\cite{Roy23} used a {\sat}-based algorithm, resulting independently to our own research in an encoding similar to the one outlined in Section \ref{sec:sat}. However, our explicit solution to the learning problem, the embedding of the negations in the syntactic DAG, the approximation of the recurrence diameter as a semantic bound, our implementation of this algorithm, its test suite, and the experimental results are entirely novel contributions.

  \section{Preliminary Definitions}
  \label{sec:def}

  \subsection{Kripke structures}
  
  Let $\ap$ be a finite set of atomic propositions. A Kripke structure is a finite directed graph whose vertices (called \emph{states}) are labelled by subsets of $\ap$.
  
  \begin{definition}[Kripke Structure]
    A \emph{Kripke structure} (KS) $\kripke$ on $\ap$ is a tuple $\kripke = (Q, \delta, \lambda)$ such that:
    \begin{itemize}
      \item $Q$ is a finite set of states; the integer $|Q|$ is known as the \emph{size} of $\kripke$;
      \item $\delta: Q \to (2^Q \setminus \{\emptyset\})$ is a transition function; the integer $\underset{q \in Q}{\max} \, |\delta(q)|$ is known as the \emph{degree} of $\kripke$;
      \item $\lambda: Q \to 2^\ap$ is a labelling function.
    \end{itemize}
  \end{definition}
  
  An infinite \emph{run} $r$ of $\kripke$ starting from a state $q \in Q$ is an infinite sequence $r = (s_i) \in Q^\omega$ of consecutive states such that $s_0 = q$ and $\forall i \geq 0$, $s_{i+1} \in \delta(s_i)$. $\runs_\kripke(q)$ is the set of all infinite runs of $\kripke$ starting from $q$.
  
  The \emph{recurrence diameter} $\rdiameter_\kripke(q)$ of state $q$ in $\kripke$ is the length of the longest finite run $(s_i)_{i = 0, \ldots, \rdiameter_\kripke(q)}$ starting from $q$ such that $\forall i, j \in [0 \isep \rdiameter_\kripke(q)]$, if $i \neq j$ then $s_i \neq s_j$ (i.e. the longest \emph{simple} path in the underlying graph structure). We may omit the index $\kripke$ whenever contextually obvious.

  Note that two states may generate the same runs despite their computation trees not corresponding. It is therefore necessary to define an equivalence relation on states of KSs that goes further than mere run equality.
   
  \begin{definition}[Bisimulation relation]
    Let $\kripke = (Q, \delta, \lambda)$ be a KS on $\ap$.
    The \emph{canonical bisimulation} relation ${\sim} \subseteq Q \times Q$ is the coarsest (i.e. the most general) equivalence relation such that for any $q_1 \sim q_2$, $\lambda(q_1) = \lambda(q_2)$ and $\forall q'_1 \in \delta(q_1), \exists q'_2 \in \delta(q_2)$ such that $q'_1 \sim q'_2$.
    \label{def:bisim}
   \end{definition}
 
   Bisimilarity does not only entails equality of runs, but also similarity of shape: two bisimilar states have corresponding computation trees at any depth. A partition refinement algorithm allows one to compute $\sim$ by refining a sequence of equivalence relations $(\sim_i)_{i \geq 0}$ on $Q \times Q$ inductively, where for every $q_1, q_2 \in Q$:
   \begin{align*}
     q_1 \sim_0 q_2 & \iff \lambda(q_1) = \lambda(q_2) \\
     q_1 \sim_{i+1} q_2 & \iff (q_1 \sim_i q_2) \land (\{[q'_1]_{\sim_i} \mid q'_1 \in \delta(q_1)\} = \{[q'_2]_{\sim_i} \mid q'_2 \in \delta(q_2)\})
   \end{align*}
 
   Where $[q]_{\sim_i}$ stands for the equivalence class of $q \in Q$ according to the equivalence relation $\sim_i$. Intuitively, $q_1 \sim_i q_2$ if their computation trees are corresponding up to depth $i$. The next theorem is a well-known result~\cite[Alg. 31]{PoMC}:
   
   \begin{theorem}[Characteristic number]
     Given a KS $\kripke$, there exists $i_0 \in \integers$ such that $\forall i \geq i_0$, ${\sim} = {\sim_i}$. The smallest integer $i_0$ verifying that property is known as the \emph{characteristic number} $\charnum_\kripke$ of $\kripke$.
   \end{theorem}
 
   Note that Browne et al.~\cite{Browne88} introduced an equivalent definition: the characteristic number of a KS is also the smallest integer $\charnum_\kripke \in \integers$ such that any two states are not bisimilar if and only if their labelled computation trees of depth $\charnum_\kripke$ are not corresponding.

  \subsection{Computation Tree Logic}
  \label{sec:ctl}
  
  \begin{definition}[Computation Tree Logic]
    \emph{Computation Tree Logic} (\ctl) is the set of formulas defined by the following grammar, where $a \in \ap$ is any atomic proposition and $\dagger \in \{\forall, \exists\}$ a quantifier:
    \begin{equation*}
      \varphi ::= a~\mid~\top~\mid~\neg\varphi~\mid~\varphi \land \varphi~\mid~\varphi \lor \varphi
        ~\mid~\dagger\nxt \varphi~\mid~\dagger\finally\varphi~\mid~\dagger \globally\varphi
        ~\mid~\dagger \, \varphi \until \varphi
    \end{equation*}
    Given $E \subseteq \{\neg, \land, \lor, \ax, \ex, \af, \ef, \ag, \eg, \au, \eu\}$, we define the syntactic fragment $\ctl(E)$ as the subset of CTL formulas featuring only operators in $E$.
  \end{definition}

  {\ctl} formulas are verified against states of KSs (a process known as \emph{model checking}). Intuitively, ${\oldforall}$ (\emph{all}) means that all runs starting from state $q$ must verify the property that follows, ${\oldexists}$ (\emph{exists}), that at least one run starting from $q$ must verify the property that follows, $\nxt \varphi$ (\emph{next}), that the next state of the run must verify $\varphi$, $\finally \varphi$ (\emph{finally}), that there exists a state of run verifying $\varphi$, $\globally \varphi$ (\emph{globally}), that each state of the run must verify $\varphi$, and $\varphi \until \psi$ (\emph{until}), that the run must keep verifying $\varphi$ at least until $\psi$ is eventually verified. 

  More formally, for a state $q\in Q$ of a KS $\kripke = (Q, \delta, \lambda)$ and a {\ctl} formula $\varphi$, we write $(q \models_\kripke \varphi)$ when $\kripke$ \emph{satisfies} $\varphi$. {\ctl}'s semantics are defined inductively on $\varphi$ (see~\cite[Def. 6.4]{PoMC} for a complete definition); we recall below the \emph{until} case:
  \begin{definition}[Semantics of ${\au,\eu}$]
    Let $\kripke = (Q, \delta, \lambda)$ be a KS, $\varphi$ and $\psi$ two
    {\ctl} formulas, $q \in Q$, and $\dagger \in \{\forall, \exists\}$. Then:
    \[q \models_\kripke \dagger \, \varphi \until \psi \iff \dagger (s_i) \in \runs_\kripke(q), \exists i \geq 0, (s_i \models_\kripke \psi) \land (\forall j < i, s_j \models_\kripke \varphi)\]
  \end{definition}
    
  Bisimilarity and {\ctl} equivalence coincide \cite[Thm. 7.20]{PoMC} on finite KSs. The proof relies on the following concept:
  
  \begin{theorem}[Browne et al.~\cite{Browne88}]
    Given a KS $\kripke = (Q, \delta, \lambda)$ and a state $q \in Q$, there exists a {\ctl} formula $\varphi_q \in \ctl(\{\neg, \land, \lor, \ax, \ex\})$ known as the \emph{master formula} of state $q$ such that, for any $q' \in Q$, $q' \models_\kripke \varphi_q$ if and only if $q \sim q'$.
    \label{th:ctl_charac}
  \end{theorem}
  

  \noindent
  \begin{minipage}{0.49\linewidth}
    \setlength{\parindent}{15pt}
    To each {\ctl} formula $\varphi$, we associate a syntactic tree $\tree$. For brevity's sake, we consider a syntactic \emph{directed acyclic graph} (DAG) $\sdag$ by coalescing identical subtrees in the original syntactic tree $\tree$, as shown in Figure \ref{fig:syntatic_dag}. The \emph{size} $|\varphi|$ of a {\ctl} formula $\varphi$ is then defined as the number of nodes of its smallest syntactic DAG. As an example, $|\neg a \land \ax a| = 4$.
  \end{minipage}
  \hspace{0.02\linewidth}
  \begin{minipage}[][][s]{0.47\linewidth}
    \begin{figure}[H]
      \centering
      \begin{tikzpicture}[x=1.5cm, y=0.75cm]
        \node (or) at (0.5,2) {$\land$};
        \node (neg) at (0,1) {$\neg$};
        \node (ax) at (1,1) {$\ax$};
        \node (a1) at (0,0) {$a$};
        \node (a2) at (1,0) {$a$};
        
        \draw[->] (or) -- (neg);
        \draw[->] (or) -- (ax);
        \draw[->] (neg) -- (a1);
        \draw[->] (ax) -- (a2);
      \end{tikzpicture}
      \hspace{0.5cm}
      \begin{tikzpicture}[x=1.5cm, y=0.75cm]
        \node (or) at (0.5,2) {$\land_{[4]}$};
        \node (neg) at (0,1) {$\neg_{[3]}$};
        \node (ax) at (1,1) {$\ax_{[2]}$};
        \node (a) at (0.5,0) {$a_{[1]}$};
        
        \draw[->] (or) -- (neg);
        \draw[->] (or) -- (ax);
        \draw[->] (neg) -- (a);
        \draw[->] (ax) -- (a);
      \end{tikzpicture}
      \caption{The syntactic tree and indexed DAG of the {\ctl} formula $\neg a \land \ax a$.}
      \label{fig:syntatic_dag}
    \end{figure}
  \end{minipage}

  \subsection{Bounded semantics}
  
  We introduce the bounded temporal operators $\af^u$, $\ef^u$, $\ag^u$, $\eg^u$, $\au^u$, and $\eu^u$, whose semantics only applies to the first $u$ steps of a run. Formally:
  
  \begin{definition}[Bounded semantics of {\ctl}]
    Let $\kripke = (Q, \delta, \lambda)$ be a KS, $\varphi$ and $\psi$ two {\ctl} formulas, $u \in \integers$ and $q \in Q$. The bounded semantics of {\ctl} of rank $u$ with regards to $\kripke$ are defined as follows for the quantifier $\dagger \in \{\forall, \exists\}$:
    \begin{align*}
      q \models_\kripke \dagger\finally^u \varphi & \iff \dagger (s_i) \in \runs_\kripke(q), \exists i \in [0 \isep u], s_i \models_\kripke \varphi \\
      q \models_\kripke \dagger\globally^u \varphi & \iff \dagger (s_i) \in \runs_\kripke(q), \forall i \in [0 \isep u], s_i \models_\kripke \varphi \\
      q \models_\kripke \dagger \, \varphi \until^u \psi & \iff \dagger (s_i) \in \runs_\kripke(q), \exists i \in [0 \isep u], (s_i \models_\kripke \psi) \land (\forall j < i, s_j \models_\kripke \varphi)
    \end{align*}
  \end{definition}

  Intuitively, the rank $u$ of the bounded semantics acts as a timer: $(q \models_\kripke \ag^u \varphi)$ means that $\varphi$ must hold for the next $u$ computation steps; $(q \models_\kripke \af^u \varphi)$, that $q$ must always be able to reach a state verifying $\varphi$ within $u$ steps; $(q \models_\kripke \forall \varphi \until^u \psi)$, that $q$ must always be able to reach a state verifying $\psi$ within $u$ steps, and that $\varphi$ must hold until it does; etc. This intuition results in the following properties:

  \begin{property}[Base case]
    $(q \models_\kripke \psi) \iff (q \models_\kripke \dagger\finally^0 \psi) \iff (q \models_\kripke \dagger \, \varphi \until^0 \psi) \\ \iff (q \models_\kripke \dagger\globally^0 \psi)$.
    \label{prop:sem_base}
  \end{property}
  
  \begin{property}[Induction]
    $(q \models_\kripke \dagger\finally^{u+1} \varphi) \iff (q \models_\kripke \varphi) \lor \underset{q' \in \delta(q)}{\bigtriangleup} (q' \models_\kripke \dagger\finally^u \varphi)$,
    $(q \models_\kripke \dagger \, \varphi \until^{u+1} \psi) \iff (q \models_\kripke \psi) \lor \biggl[(q \models_\kripke \varphi) \land \underset{q' \in \delta(q)}{\bigtriangleup} (q' \models_\kripke \dagger \, \varphi \until^u \psi)\biggr]$,
    and $(q \models_\kripke \dagger\globally^{u+1} \varphi) \iff (q \models_\kripke \varphi) \land \underset{q' \in \delta(q)}{\bigtriangleup} (q' \models_\kripke \dagger\globally^u \varphi)$, where ${\vartriangle} = \land$ if $\dagger = \oldforall$ and ${\vartriangle} = \lor$ if $\dagger = \oldexists$.
    \label{prop:sem_induction}
  \end{property}
  
  \begin{property}[Spread]
    $(q \models_\kripke \dagger\finally^u \varphi) \implies (q \models_\kripke \dagger\finally^{u+1} \varphi)$, $(q \models_\kripke \dagger\globally^{u+1} \varphi) \implies (q \models_\kripke \dagger\globally^u \varphi)$, and $(q \models_\kripke \dagger \, \varphi \until^u \psi) \implies (q \models_\kripke \dagger \, \varphi \until^{u+1} \psi)$.
    \label{prop:sem_spread}
  \end{property}
  
  Bounded model checking algorithms~\cite{Zhang09} rely on the following result, as one can then restrict the study of {\ctl} semantics to finite and fixed length paths.

  \begin{theorem}
    Given $q \in Q$, for $\dagger \in \{\forall, \exists\}$ and $\genop \in \{\finally, \globally\}$, $q \models_\kripke \dagger\genop \varphi$ (resp. $q \models_\kripke \dagger \, \varphi \until \psi$) if and only if $q \models_\kripke \dagger\triangleright^{\rdiameter(q)} \varphi$ (resp. $q \models_\kripke \dagger \, \varphi \until^{\rdiameter(q)} \psi$).
    \label{th:bounded_sem}
  \end{theorem}

  A full proof of this result is available in Appendix \ref{app:proof_bmc}.

  \section{The Learning Problem}
  \label{sec:learning}
  
  We consider the synthesis problem of a distinguishing {\ctl} formula from a sample of positive and negative states of a given KS.

  \subsection{Introducing the problem}
  
  First and foremost, the sample must be self-consistent: a state in the positive sample cannot verify a {\ctl} formula while another bisimilar state in the negative sample does not.
  
  \begin{definition}[Sample]
    Given a KS $\kripke = (Q, \delta, \lambda)$, a \emph{sample} of $\kripke$ is a pair $(\sample) \in 2^Q \times 2^Q$ such that $\forall q^+ \in \psample$, $\forall q^- \in \nsample$, $q^+ \not\sim q^-$.
  \end{definition}

  We define the \emph{characteristic number} $\charnum_\kripke(\sample)$ of a sample as the smallest integer $c \in \integers$ such that for every $q^+ \in \psample$, $q^- \in \snsample$, $q^+ \not\sim_c q^-$.
  
  \begin{definition}[Consistent formula]
    A {\ctl} formula $\varphi$ is said to be \emph{consistent} with a sample $(\sample)$ of $\kripke$ if $\forall q^+ \in \psample$, $q^+ \models_\kripke \varphi$ and $\forall q^- \in \nsample$, $q^- \not\models_\kripke \varphi$.
  \end{definition}

  The rest of our article focuses on the following passive learning problems:
  
  \begin{definition}[Learning problem]
    Given a sample $(\sample)$ of a KS $\kripke$ and $n \in \integers^*$, we introduce the following instances of the {\ctl} learning problem:
    \begin{description}
      \item[$\learn_{\ctl(E)}(\kripke, \sample)$.] Is there $\varphi \in \ctl(E)$ consistent with $(\sample)$?
      \item[$\learn_{\ctl(E)}^{\leq n}(\kripke, \sample)$.] Is there $\varphi \in \ctl(E)$, $|\varphi| \leq n$, consistent with $(\sample)$?
      \item[$\minlearn_{\ctl(E)}(\kripke, \sample)$.] Find the \emph{smallest} $\varphi \in \ctl(E)$ consistent with $(\sample)$.
    \end{description}
    \label{def:learning}
  \end{definition}
  
  \begin{theorem}
    $\learn_{\ctl}(\kripke, \sample)$ and $\minlearn_{\ctl}(\kripke, \sample)$ always admit a solution.
    \label{th:solution}
  \end{theorem}

  \begin{proof}
    Consider $\psi = \underset{q^+ \in \psample}{\bigvee} \varphi_{q^+}$. This formula $\psi$ is consistent with $(\psample, \nsample)$ by design. Thus $\learn_{\ctl}(\kripke, \sample)$ always admits a solution, and so does the problem $\minlearn_{\ctl}(\kripke, \sample)$, although $\psi$ is unlikely to be the minimal solution. \hfill\qed
  \end{proof}
  
  \smallskip
  Bordais et al.~\cite{Bordais2023} proved that $\learn_{\ctl}^{\leq n}(\kripke, \sample)$ is {\np}-hard, assuming the set of atomic propositions $\ap$ is given as an input as well.

  \subsection{An explicit solution}

  We must find a formula consistent with the sample $(\sample)$, an easier problem than Browne et al.~\cite{Browne88}'s answer to Theorem \ref{th:ctl_charac} that subsumes bisimilarity with an entire KS. As we know that every state in $\nsample$ is dissimilar to every state in $\psample$, we will try to encode this fact in $\ctl$ form, then use said encoding to design a formula consistent with the sample.
  
  \begin{definition}[Separating formula]
    Let $(\sample)$ be a sample of a KS $\kripke = (Q, \delta, \lambda)$. Assuming that $\ap$ and $Q$ are ordered, and given $q_1, q_2 \in Q$ such that $q_1 \not\sim q_2$, formula $D_{q_1, q_2}$ is defined inductively w.r.t. $c = \charnum_\kripke(\{q_1\}, \{q_2\})$ as follows:
    \begin{itemize}
      \item if $c = 0$ and $\lambda(q_1) \setminus \lambda(q_2) \neq \emptyset$ has minimal element $a$, then $D_{q_1, q_2} = a$;
      \item else if $c = 0$ and $\lambda(q_2) \setminus \lambda(q_1) \neq \emptyset$ has minimal element $a$, then $D_{q_1, q_2} = \neg a$;
      \item else if $c \neq 0$ and $\exists q_1' \in \delta(q_1)$, $\forall q_2' \in \delta(q_2)$, $q_1' \not\sim_{c-1} q_2'$, then $D_{q_1, q_2} = \ex \, \biggl( \underset{q_2' \in \delta(q_2)}{\bigwedge} D_{q_1', q'_2} \biggr)$, picking the smallest $q_1'$ verifying this property;
      \item else if $c \neq 0$ and $\exists q_2' \in \delta(q_2)$, $\forall q_1' \in \delta(q_1)$, $q_1' \not\sim_{c-1} q_2'$, then $D_{q_1, q_2} = \ax \, \neg \biggl( \underset{q_1' \in \delta(q_1)}{\bigwedge} D_{q_2', q_1'} \biggr)$, picking the smallest $q_2'$ verifying this property.
    \end{itemize}
    The formula $\sepfor_\kripke(\sample) = \underset{q^+ \in \spsample}{\bigvee} \underset{q^- \in \snsample}{\bigwedge} D_{q^+, q^-} \in \ctl(\{\neg, \land, \lor, \ax, \ex\})$ is then called the \emph{separating formula} of sample $(\sample)$.
  \end{definition}
  
  Intuitively, the $\ctl$ formula $D_{q_1, q_2}$ merely expresses that states $q_1$ and $q_2$ are dissimilar by negating Definition \ref{def:bisim}; it is such that $q_1 \models_\kripke D_{q_1, q_2}$ but $q_2 \not\models_\kripke D_{q_1, q_2}$. Either $q_1$ and $q_2$ have different labels, $q_1$ admits a successor that is dissimilar to $q_2$'s successors, or $q_2$ admits a successor that is dissimilar to $q_1$'s. The following result is proven in Appendix \ref{app:proof_explicit_sol}:
  
  \begin{theorem}
    The separating formula $\sepfor_\kripke(\sample)$ is consistent with $(\sample)$.
    \label{th:explicit_sol}
  \end{theorem}

  As proven in Appendix \ref{app:proof_size_explicit}, we can bound the size of $\sepfor_\kripke(\sample)$:

  \begin{corollary}
    Assume the KS $\kripke$ has degree $k$ and $c = \charnum_\kripke(S^+, S^-)$, then:
    \begin{itemize}
      \item if $k \geq 2$, then $|\sepfor_\kripke(\sample)| \leq (5 \cdot k^c + 1) \cdot |\psample| \cdot |\nsample|$;
      \item if $k = 1$, then $|\sepfor_\kripke(\sample)| \leq (2 \cdot c + 3) \cdot |\psample| \cdot |\nsample|$.
    \end{itemize}
    \label{cor:size_explicit_sol}
  \end{corollary}

  \section{SAT-based Learning}
  \label{sec:sat}

  The \emph{universal} fragment $\ctla = \ctl(\{\neg, \land, \lor, \ax, \af, \ag, \au\}$) of {\ctl} happens to be as expressive as the full logic~\cite[Def. 6.13]{PoMC}. For that reason, we will reduce a learning instance of $\ctla$ of rank $n$ to an instance of the {\sat} problem. A similar reduction has been independently found by Roy et al.~\cite{Roy23}.
  
  \begin{lemma}
    There exists a Boolean propositional formula $\Phi_n$ such that the instance $\learn_\ctla^{\leq n}(\kripke, \spsample, \snsample)$ of the learning problem admits a solution $\varphi$ if and only if the formula $\Phi_n$ is satisfiable.
    \label{lem:exact}
  \end{lemma}

  \subsection{Modelling the formula}
  \label{subec:sat_dag}
  
  Assume that there exists a syntactic DAG $\sdag$ of size smaller than or equal to $n$ representing the desired $\ctl$ formula $\varphi$. Let us index $\sdag$'s nodes in $[1 \isep n]$ in such a fashion that each node has a higher index than its children, as shown in Figure \ref{fig:syntatic_dag}. Hence, $n$ always labels a root and $1$ always labels a leaf.
  
  Let $\labels = \ap \cup \{\top, \neg, \land, \lor, \ax, \af, \ag, \au\}$ be the set of labels that decorates the DAG's nodes. For each $i\ \in [1 \isep n]$ and $o \in \labels$, we introduce a Boolean variable $\tau_i^o$ such that $\tau_i^o = 1$ if and only if the node of index $i$ is labelled by $o$.
  
  For all $i \in [1 \isep n]$ and $j \in [0 \isep i-1]$, we also introduce a Boolean variable $\leftv_{i, j}$ (resp. $\rightv_{i, j}$) such that $\leftv_{i, j} = 1$ (resp. $\rightv_{i, j} = 1$) if and only if $j$ is the left (resp. right) child of $i$. Having a child of index $0$ stands for having no child at all in the actual syntactic DAG $\sdag$.
  
  Three mutual exclusion clauses \ref{sat:tree:unique_label}, \ref{sat:tree:unique_left}, and \ref{sat:tree:unique_right} guarantee that each node of the syntactic DAG has exactly one label and at most one left child and one right child. Moreover, three other clauses \ref{sat:tree:leaf_no_children}, \ref{sat:tree:unary_one_child}, and \ref{sat:tree:binary_two_children} ensure that a node labelled by an operator of arity $x$ has exactly $x$ actual children (by convention, if $x = 1$ then its child is to the right). These simple clauses are similar to Neider's encoding~\cite{Neider18} and for that reason are only detailed in Appendix \ref{app:sat_tree}.

  \subsection{Applying the formula to the sample}
  \label{subsec:sat_semantics}
    
  For all $i \in [1 \isep, n]$ and $q \in Q$, we introduce a Boolean variable $\varphi_i^q$ such that $\varphi_i^q = 1$ if and only if state $q$ verifies the sub-formula $\varphi_i$ rooted in node $i$. The next clauses implement the semantics of the true symbol $\top$, the atomic propositions, and the {\ctl} operator $\ax$.
  {\allowdisplaybreaks
  \small
  \begin{align}
    \underset{\substack{i \in [1 \isep n] \\ q \in Q}}{\bigwedge} &
    (\tau_i^\top \implies \varphi_i^{q})
    \tag{$\mathtt{sem}_\top$}
    \label{sat:system:top} \\
    \underset{\substack{i \in [1 \isep n] \\ q \in Q}}{\bigwedge} & \Biggl[
    \biggl(\underset{a \in \lambda(q)}{\bigwedge} (\tau_i^a \implies \varphi_i^{q}) \biggr)
    \land \biggl(\underset{a \not\in \lambda(q)}{\bigwedge} (\tau_i^a \implies \neg \varphi_i^{q}) \biggr)
    \Biggr]
    \tag{$\mathtt{sem}_a$}
    \label{sat:system:predicates} \\
    \underset{\substack{i \in [2 \isep n] \\ k \in [1 \isep i-1]}}{\bigwedge} \Biggl[
    & (\tau_i^{\ax} \land \rightv_{i, k}) \implies
    \underset{q \in Q}{\bigwedge} \Biggl(\varphi_i^{q} \iff \underset{q' \in \delta(q)}{\bigwedge} \varphi_k^{q'}\Biggr)
    \Biggr]
    \tag{$\mathtt{sem}_{\ax}$}
    \label{sat:system:all_next}
  \end{align}}

  Semantic clauses are structured as follows: an antecedent stating node $i$'s label and its possible children implies a consequent expressing $\varphi_i^{q}$'s semantics for each $q \in Q$. Clause \ref{sat:system:top} states that $q \models \top$ is always true; clause \ref{sat:system:predicates}, that $q \models a$ if and only if $q$ is labelled by $a$; and clause \ref{sat:system:all_next}, that $q \models \ax \psi$ if and only if all of $q$'s successors verify $\psi$. Similar clauses \ref{sat:system:not}, \ref{sat:system:and}, and \ref{sat:system:or} encoding the semantics of the Boolean connectors $\neg$, $\land$ and $\lor$ are only detailed in Appendix \ref{app:sat_tree}, being straightforward enough.
      
  For reasons detailed in Appendix \ref{app:wrong_semantics}, we will encode the bounded semantics of the temporal operators $\af^u$, $\ag^u$ and $\au^u$. For all $i \in [1 \isep n]$, $q \in Q$, and $u \in [0 \isep \alpha(q)]$, we introduce a Boolean variable $\rank_{i,q}^u$ such that $\rank_{i,q}^u = 1$ if and only if $q$ verifies the sub-formula rooted in $i$ according to the {\ctl} bounded semantics of rank $u$ (e.g. $q \models \af^u \psi$, assuming sub-formula $\af \psi$ is rooted in node $i$).
  
  Thanks to Theorem \ref{th:bounded_sem} we can introduce the following equivalence clause:
  {\allowdisplaybreaks
  \small
  \begin{align}
    \underset{i \in [2 \isep n]}{\bigwedge} & \Biggl[
    \Biggl(\underset{o \in \{\af, \ag, \au\}}{\bigvee} \tau_i^o\Biggr) \implies
    \underset{q \in Q}{\bigwedge} (\varphi_i^q \iff\rank_{i,q}^{\rdiameter(q)})
    \Biggr]
    \tag{$\mathtt{sem}_{\rank}$}
    \label{sat:system:rank_semantics}
  \end{align}}
 
  Property \ref{prop:sem_spread} yields two other clauses whose inclusion is not mandatory (they were left out by Roy et al.~\cite{Roy23}) that further constrains the bounded semantics:
  {\allowdisplaybreaks
  \small
  \begin{align}
    \underset{i \in [2 \isep n]}{\bigwedge} &
    \Biggl[(\tau_i^{\af} \lor \tau_i^{\au}) \implies 
    \underset{\substack{q \in Q \\ u \in [1 \isep \rdiameter(q)]}}{\bigwedge} (\rank_{i,q}^{u-1} \implies \rank_{i,q}^{u})
    \Biggr]
    \tag{$\mathtt{ascent}_{\rank}$}
    \label{sat:system:rank_ascent} \\
    \underset{i \in [2 \isep n]}{\bigwedge} &
    \Biggl[\tau_i^{\ag} \implies 
    \underset{\substack{q \in Q \\ u \in [1 \isep \rdiameter(q)]}}{\bigwedge} (\rank_{i,q}^{u} \implies \rank_{i,q}^{u-1})
    \Biggr]
    \tag{$\mathtt{descent}_{\rank}$}
    \label{sat:system:rank_descent}
  \end{align}}
  
  The next clause enables variable $\rank_{i,q}^u$ for temporal operators only:
  {\allowdisplaybreaks
  \small
  \begin{align}
    \underset{i \in [2 \isep n]}{\bigwedge} &
    \Biggl[\Biggl(\underset{o \in \{\af, \ag, \au\}}{\bigwedge} \neg \tau_i^o\Biggr) \implies
    \underset{\substack{q \in Q \\ u \in [0 \isep \rdiameter(q)]}}{\bigwedge} \neg\rank_{i,q}^u
    \Biggr]
    \tag{$\mathtt{no}_{\rank}$}
    \label{sat:system:rank_forbidden}
  \end{align}}

  Properties \ref{prop:sem_base} and \ref{prop:sem_induction} yield an inductive definition of bounded semantics. We only explicit the base case \ref{sat:system:rank_start} and the semantics \ref{sat:system:all_until} of $\au^u$, but also implement clauses \ref{sat:system:all_finally} and \ref{sat:system:all_globally} detailed in Appendix \ref{app:sat_universal_ops}.
  {\allowdisplaybreaks
  \small
  \begin{align}
    \underset{\substack{i \in [2 \isep n] \\ k \in [1 \isep i-1]}}{\bigwedge} & \Biggl[
    \Biggl(\Biggl[\underset{o \in \{\af, \ag, \au\}}{\bigvee} \tau_i^o\Biggr] \land \rightv_{i, k}\Biggr) \implies
    \underset{q \in Q}{\bigwedge} (\rank_{i,q}^0 \iff \varphi_k^q)
    \Biggr]
    \tag{$\mathtt{base}_{\rank}$}
    \label{sat:system:rank_start} \\
    \underset{\substack{i \in [2 \isep n] \\ j, k \in [1 \isep i-1]}}{\bigwedge} & \Biggl[
    (\tau_i^{\au} \land \leftv_{i, j} \land \rightv_{i, k}) \implies \nonumber \\
    & \underset{\substack{q \in Q \\ u \in [1 \isep \rdiameter(q)]}}{\bigwedge} \Biggl(\rank_{i,q}^u \iff \Biggl[\varphi_k^q \lor \Biggl(\varphi_j^q \land \underset{q' \in \delta(q)}{\bigwedge} \rank_{i, q'}^{\min(\rdiameter(q'), u-1)}\Biggr)\Biggr]\Biggr)
    \Biggr]
    \tag{$\mathtt{sem}_{\au}$}
    \label{sat:system:all_until}
  \end{align}}

  Finally, the last clause ensures that the full formula $\varphi$ (rooted in node $n$) is verified by the positive sample but not by the negative sample.
  {\small
  \begin{align}
    \Biggl(\underset{{q^+} \in \spsample}{\bigwedge} \varphi_n^{q^+} \Biggr)
    \land \Biggl(\underset{q^- \in \snsample}{\bigwedge} \neg \varphi_n^{q^-} \Biggr)
    \tag{$\mathtt{sem}_\varphi$}
    \label{sat:system:sample}
  \end{align}}

  \subsection{Solving the SAT instance}
  
  We finally define the formula $\Phi_n$ as the conjunction of all the aforementioned clauses. Assuming an upper bound $d$ on the KS's recurrence diameter, this encoding requires $\bigo(n^2 + n \cdot |\ap| + n \cdot |Q| \cdot d)$ variables and $\bigo(n \cdot |\ap| + n^3 \cdot |Q| \cdot d + n \cdot |\ap| \cdot |Q|)$ clauses, not taking transformation to conjunctive normal form into account. By design, Lemma \ref{lem:exact} holds.

  \begin{proof}
    Clauses \ref{sat:tree:unique_label} to \ref{sat:tree:binary_two_children} allow one to infer the DAG of a formula $\varphi \in \ctl$ of size smaller than or equal to $n$ from the valuations taken by the variables $(\tau_i^o)$, $(\leftv_{i, j})$, and $(\rightv_{i, j})$. Clauses \ref{sat:system:predicates} to \ref{sat:system:sample} guarantee that the sample is consistent with said formula $\varphi$, thanks to Theorem \ref{th:bounded_sem} and Properties \ref{prop:sem_base}, \ref{prop:sem_induction}, and \ref{prop:sem_spread}. \hfill\qed
  \end{proof}

  \section{Algorithms for the Minimal Learning Problem}
  \label{sec:algo}
  
  We introduce in this section an algorithm to solve the minimum learning problem $\minlearn_\ctla(\kripke, \psample, \nsample)$. Remember that it always admits a solution if and only if the state sample is consistent by Theorem \ref{th:solution}.

  \subsection{A bottom-up algorithm}
  
  {\centering
  \begin{minipage}{0.46\linewidth}
    By Theorem \ref{th:solution}, there exists a rank $n_0$ such that the problem $\learn_{\ctla}^{\leq n_0}(\kripke, \ssample)$ admits a solution. Starting from $n = 0$, we can therefore try to solve $\learn_\ctla^{\leq n}(\kripke, \ssample)$ incrementally until a (minimal) solution is found, in a similar manner to Neider and Gavran~\cite{Neider18}. Algorithm \ref{alg:iter_learn} terminates with an upper bound $n_0$ on the number of required iterations.
  \end{minipage}
  \hfill
  \begin{minipage}{0.53\linewidth}
    \begin{algorithm}[H]\small
      \caption{Solving $\minlearn_\ctla(\kripke, \ssample)$.}
      \label{alg:iter_learn}
      \KwIn{a KS $\kripke$ and a sample $\ssample$.}
      \KwOut{the smallest $\ctla$ formula $\varphi$ consistent with $\ssample$.}
      $n \gets 0$\;
      \Repeat{$\Phi_n$ is satisfiable by some valuation $v$}{
        $n \gets n+1$\;
        compute $\Phi_n$\;
      }
      from $v$ build and \Return{$\varphi$}.
      \medskip
    \end{algorithm}
  \end{minipage}}

  \subsection{Embedding negations}
  \label{sec:embed_negation}
  
  The {\ctl} formula $\ef a$ is equivalent to the $\ctla$ formula $\neg \ag \neg a$, yet the former remains more succinct, being of size $2$ instead of $4$. While $\ctla$ has been proven to be as expressive as {\ctl}, the sheer amount of negations needed to express an equivalent formula can significantly burden the syntactic DAG. A possible optimization is  to no longer consider the negation $\neg$ as an independent operator but instead embed it in the nodes of the syntactic DAG, as shown in Figure \ref{fig:neg_dag}.
  
  \begin{figure}
    \centering
    \begin{tikzpicture}[x=1cm, y=0.4cm]
      \node (neg1) at (0,0) {$\neg$};
      \node (eu) at (1,0) {$\eu$};
      \node (top) at (2,-1) {$\top$};
      \node (neg2) at (2,1) {$\neg$};
      \node (ax) at (3,1) {$\ax$};
      \node (neg3) at (4,1) {$\neg$};
      \node (a) at (5,1) {$a$};
      
      \draw[->] (neg1) -- (eu);
      \draw[->] (eu) -- (top);
      \draw[->] (eu) -- (neg2);
      \draw[->] (neg2) -- (ax);
      \draw[->] (ax) -- (neg3);
      \draw[->] (neg3) -- (a);
    \end{tikzpicture}
    \hspace{1cm}
    \begin{tikzpicture}[x=1.5cm, y=0.4cm]
      \node (eu) at (0,0) {$\neg \eu_{[1]}$};
      \node (top) at (1,-1) {$\top_{[2]}$};
      \node (ax) at (1,1) {$\neg \ax_{[3]}$};
      \node (a) at (2,1) {$\neg a_{[4]}$};
      
      \draw[->] (eu) -- (top);
      \draw[->] (eu) -- (ax);
      \draw[->] (ax) -- (a);
    \end{tikzpicture}
    \caption{The syntactic DAG of $\neg \exists \top \until \neg \ax \neg a$, before and after embedding negations.}
    \label{fig:neg_dag}
  \end{figure}

  Note that such a definition of the syntactic DAG alters one's interpretation of a {\ctl} formula's size: as a consequence, under this optimization, Algorithm \ref{alg:iter_learn} may yield a formula with many negations that is no longer minimal under the original definition of size outlined in Section \ref{sec:ctl}.

  Formally, for each $i \in [1 \isep n]$, we introduce a new variable $\nu_i$ such that $\nu_i = 0$ if and only if the node of index $i$ is negated. As an example, in Figure \ref{fig:neg_dag}, $\nu_1 = \nu_3 = \nu_4 = 0$, but $\nu_2 = 1$ and the sub-formula rooted in node $3$ is $\neg \ax \neg a$.
    
  We then change the {\sat} encoding of $\ctla$'s semantics accordingly. We remove the $\neg$ operator from clauses \ref{sat:tree:unique_label}, \ref{sat:tree:unary_one_child}, and the set $\labels$ of labels. We delete clause \ref{sat:system:not}. Finally, we update the semantic clauses \ref{sat:system:top}, \ref{sat:system:predicates}, \ref{sat:system:and}, \ref{sat:system:or}, \ref{sat:system:all_next}, and \ref{sat:system:rank_semantics}. Indeed, the right side of each equivalence expresses the semantics of the operator rooted in node $i$ \emph{before} applying the embedded negation; we must therefore change the left side of the semantic equivalence accordingly, replacing the Boolean variable $\varphi_i^q$ with the formula $\tilde{\varphi}_i^q = (\neg \nu_i \land \neg\varphi_i^q) \lor (\nu_i \land \varphi_i^q)$ that is equivalent to $\varphi_i^q$ if $\nu_i = 1$ and $\neg \varphi_i^q$ if $\nu_i = 0$. These changes are fully described in Appendix \ref{app:embedded_neg}.

  \subsection{Optimizations and alternatives}
  
  \paragraph{Minimizing the input KS.}
  
  In order to guarantee that an input $\ssample$ is indeed a valid sample, one has to ensure no state in the positive sample is bisimilar to a state in the negative sample. To do so, one has to at least partially compute the bisimilarity relation $\sim$ on $\kripke = (Q, \delta, \lambda)$. But refining it to completion can be efficiently performed in $\bigo(|Q| \cdot |\ap| + |\delta| \cdot \log(|Q|))$ operations~\cite[Thm. 7.41]{PoMC}, yielding a bisimilar KS $\kripke_{\min}$ of minimal size.
  
  Minimizing the input KS is advantageous as the size of the semantic clauses depends on the size of $\kripke$, and the {\sat} solving step is likely to be the computational bottleneck. As a consequence, we always fully compute the bisimulation relation $\sim$ on $\kripke$ and minimize said KS.
  
  \paragraph{Approximating the recurrence diameter.}
  \label{sec:approx_acyclic}
  
  Computing the recurrence diameter of a state $q$ is unfortunately an \np-hard problem that is known to be hard to approximate \cite{Bjorklund04}. A coarse upper bound is $\alpha(q) \leq |Q|-1$: it may however result in a significant number of unnecessary variables and clauses. Fortunately, the decomposition of a KS $\kripke$ into strongly connected components (SCCs) yields a finer over-approximation shown in Figure \ref{fig:approx_acyclic} that relies on the ease of computing $\alpha$ in a DAG. It is also more generic and suitable to {\ctl} than existing approximations dedicated to {\ltl} bounded model checking~\cite{Kroening03}.
  
  Contracting each SCC to a single vertex yields a DAG known as the \emph{condensation} of $\kripke$. We weight each vertex of this DAG $\contraction$ with the number of vertices in the matching SCC. Then, to each state $q$ in the original KS $\kripke$, we associate the weight $\beta(q)$ of the longest path in the DAG $\contraction$ starting from $q$'s SCC, minus one (in order not to count $q$). Intuitively, our approximation assumes that a simple path entering a SCC can always visit every single one of its states once before exiting, a property that obviously does not hold for two of the SCCs shown here.
  
  \begin{figure}
    \centering
    \begin{tikzpicture}[x=1.25cm, y=0.5cm]
      \node (k) at (3,-2) {$\kripke$};
      \node[state, label={90:$\mathbf{q}_0$}, minimum size=0pt] (q0) at (2,2) {};
      \node[state, label={90:$\mathbf{q}_1$}, minimum size=0pt] (q1) at (1,1) {};
      \node[state, label={270:$\mathbf{q}_2$}, minimum size=0pt] (q2) at (1,-1) {};
      \node[state, label={90:$\mathbf{q}_3$}, minimum size=0pt] (q3) at (0,0) {};
      \node[state, label={180:$\mathbf{q}_4$}, minimum size=0pt] (q4) at (2,0) {};
      \node[state, label={90:$\mathbf{q}_5$}, minimum size=0pt] (q5) at (3,0) {};
      \node[state, label={180:$\mathbf{q}_6$}, minimum size=0pt] (q6) at (2,-2) {};
      \draw[->] (q0) to [bend left] (q1);
      \draw[->] (q0) -- (q4);
      \draw[->] (q1) -- (q2);
      \draw[->] (q2) -- (q3);
      \draw[->] (q3) -- (q1);
      \draw[->] (q4) to [bend right] (q5);
      \draw[->] (q5) to [bend right] (q4);
      \draw[->] (q2) to [bend left] (q6);
      \draw[->] (q4) -- (q6);
      \draw[->] (q6) to [loop right] (q6);

      \node[draw, dotted, fit=(q0)] {};
      \node[draw, dotted, fit=(q1) (q2) (q3)] {};
      \node[draw, dotted, fit=(q4) (q5)] {};
      \node[draw, dotted, fit=(q6)] {};
      
      \node (g) at (4,2) {$\contraction$};
      \node[state, rectangle, dotted, minimum size=0pt] (c0) at (5,2) {$\mathbf{1}$};
      \node[state, rectangle, dotted, minimum size=0pt] (c1) at (4,0) {$\mathbf{3}$};
      \node[state, rectangle, dotted, minimum size=0pt] (c2) at (5,0) {$\mathbf{2}$};
      \node[state, rectangle, dotted, minimum size=0pt] (c3) at (5,-2) {$\mathbf{1}$};
      \draw[->] (c0) -- (c1);
      \draw[->] (c0) -- (c2);
      \draw[->] (c1) -- (c3);
      \draw[->] (c2) -- (c3);

      \node (alpha) at (6.5,0) [align=left]{$\alpha(q_0) = 3$ \\ $\alpha(q_1) = 2$ \\ $\alpha(q_2) = 1$ \\ $\alpha(q_3) = 3$ \\ $\alpha(q_4) = 1$ \\ $\alpha(q_5) = 2$ \\ $\alpha(q_6) = 0$};
      \node (beta) at (8,0) [align=left]{$\beta(q_0) = 4$ \\ $\beta(q_1) = 3$ \\ $\beta(q_2) = 3$ \\ $\beta(q_3) = 3$ \\ $\beta(q_4) = 2$ \\ $\beta(q_5) = 2$ \\ $\beta(q_6) = 0$};
    \end{tikzpicture}
    \caption{An approximation $\beta$ of the recurrence diameter $\alpha$ relying on SCC decomposition that improves upon the coarse upper bound $\alpha(q) \leq |Q|-1 = 6$.}
    \label{fig:approx_acyclic}
  \end{figure}
  
  \paragraph{Encoding the full logic.}
  
  $\ctla$ is semantically exhaustive but the existential temporal operators commonly appear in the literature; we can therefore consider the learning problem on the full {\ctl} logic by integrating the operators $\ex$, $\ef$, $\eg$, and $\eu$, whose Boolean encoding is discussed in Appendix \ref{app:sat_existential_ops}. We also consider the fragment $\ctlu = \{\neg, \lor, \ex, \eg, \eu\}$ used by Roy et al.~\cite{Roy23}.

  \section{Experimental Implementation}
  \label{sec:implem}
  
  We implement our learning algorithm in a \texttt{C++} prototype tool \texttt{LearnCTL}\footnote{publicly available at https://gitlab.lre.epita.fr/adrien/learnctl}, relying on Microsoft's \texttt{Z3} due to its convenient \texttt{C++} API. It takes as an input a sample of positive and negative KSs with initial states, then coalesced into a single KS and a sample of states compatible with the theoretical framework we described. It finally returns a separating $\ctla$, {\ctl}, or $\ctlu$ formula after a sanity check performed by model-checking the input KSs against the learnt formula, using a simple algorithm based on Theorem \ref{th:bounded_sem}.

  \subsection{Benchmark Collection}
  
  We intend on using our tool to generate formulas that can explain flaws in faulty implementations of known protocols. To do so, we consider structures generated by higher formalisms such as program graphs: a single mutation in the program graph results in several changes in the resulting KS. This process has been achieved manually according to the following criteria:
  \begin{itemize}
    \item The mutations only consist in deleting lines.
    \item The resulting KS should be \emph{small}, less than $\sim 1000$ states.
    \item Any mutation should result in a syntactically correct model.
  \end{itemize}
 
  We collected program graphs in a toy specification language for a CTL model checker class implemented in Java. Furthermore, we also considered PROMELA models from the \texttt{Spin} model-checker~\cite{Holzmann04} repository. Translations were then performed through the Python interface of \texttt{spot/ltsmin}~\cite{Duret22,Kant15}.
 
  \begin{example}
    \label{ex:peterson}
    Consider the mutual exclusion protocol proposed by~\cite{Peterson81} and specified in PROMELA in Figure~\ref{fig:peterson} that generates a KS with 55 states. We generate mutants by deleting no more than one line of code at a time, ignoring variable and process declarations as they are necessary for the model to be compiled and the two assertion lines that are discarded by our KS generator, our reasoning being that subtle changes yield richer distinguishing formulas.

    Furthermore, removing the instruction \texttt{ncrit}\verb|--| alone would lead to an infinite state space; thus, its deletion is only considered together with the instruction \texttt{ncrit++}. Finally, we set some atomic propositions of interest: $c$ stands for at least one process being in the critical section (\texttt{ncrit>0}), $m$ for both processes (\texttt{ncrit>1}), and $t$ for process $0$'s turn. An extra $\textit{dead}$ atomic proposition is added by Spot/LTSMin to represent deadlocked states.
    
    As summarized on Figure~\ref{fig:peterson}, every mutated model, once compared with the original KS, lead to distinguishing formulas characterizing
    Peterson's protocol: mutations \texttt{m1}, \texttt{m2}, and \texttt{m3} yield a mutual exclusion property, \texttt{m4} yields a liveness property, \texttt{m5} yields a fairness property, and \texttt{m6} yields global liveness formula.
  \end{example}

  \begin{figure}
    \begin{lstlisting}[language=promela,escapechar=!, frame=single]
!\tikz[overlay, remember picture,cm1/.style={orange},cm2/.style={red},cm3/.style={blue},cm4/.style={olive},cm5/.style={violet},cm6/.style={brown}]{
      %\node [fill=pink] (all) at ([xshift=\textwidth]pic cs:x1) {Device A};
      %\node [fill=pink] at (pic cs:m1a) {m1};
      %\fill [pink] (pic cs:m1a) rectangle (pic cs:m1b);
    
      % Highlight lines
      %  \begin{scope}[mut/.style={thick, transform canvas={yshift=-0.09em}}]
      %   \draw[mut, cm1] (pic cs:m1a) -- (pic cs:m1b);
      %   \draw[mut, cm2] (pic cs:m2a) -- (pic cs:m2b);
      %   \draw[mut, cm3] (pic cs:m3a) -- (pic cs:m3b);
      %   \draw[mut, cm4] (pic cs:m4a) -- (pic cs:m4b);
      %   \draw[mut, cm4] (pic cs:m4c) -- (pic cs:m4d);
      %   \draw[mut, cm5] (pic cs:m5a) -- (pic cs:m5b);
      %   \draw[mut, cm6] (pic cs:m6a) -- (pic cs:m6b);
      %  \end{scope}
    
      % Highlights bis
      \begin{scope}[mut/.style={thick, transform canvas={yshift=-0.24em}, opacity=0.5}]     
        \draw[rectangle, mut, cm1,fill, rounded corners=0.2em] (pic cs:m1a) rectangle ($(pic cs:m1b)+(0,1em)$);
        \draw[rectangle, mut, cm2,fill, rounded corners=0.2em] (pic cs:m2a) rectangle ($(pic cs:m2b)+(0,1em)$);
        \draw[rectangle, mut, cm3,fill, rounded corners=0.2em] (pic cs:m3a) rectangle ($(pic cs:m3b)+(0,1em)$);
        \draw[rectangle, mut, cm4,fill, rounded corners=0.2em] (pic cs:m4a) rectangle ($(pic cs:m4b)+(0,1em)$);
        \draw[rectangle, mut, cm4,fill, rounded corners=0.2em] (pic cs:m4c) rectangle ($(pic cs:m4d)+(0,1em)$);
        \draw[rectangle, mut, cm5,fill, rounded corners=0.2em] (pic cs:m5a) rectangle ($(pic cs:m5b)+(0,1em)$);
        \draw[rectangle, mut, cm6,fill, rounded corners=0.2em] (pic cs:m6a) rectangle ($(pic cs:m6b)+(0,1em)$);
      \end{scope}
    
     % Mutations:
      \begin{scope}[every node/.style={transform canvas={yshift=0.4em, xshift=1em}, thick}]
        \node[cm1] (mm1) at (pic cs:m1b) {m1};
        \node[cm2] (mm2) at (pic cs:m2b) {m2};
        \node[cm3] (mm3) at (pic cs:m3b) {m3};
        \node[cm4] (mm4) at (pic cs:m4b) {m4};
        \node[cm5] (mm5) at (pic cs:m5b) {m5};
        \node[cm6] (mm6) at (pic cs:m6b) {m6};
      \end{scope}
    
      % Distinguishing formulae:
      \begin{scope}[every node/.style={draw, rounded corners=0.3em, anchor=east}]
        \path let \p1 = (pic cs:m1b) in let \p2 = (pic cs:vert) in node (d1) at (\x2,\y1) {$\ag \neg m$};
        \path let \p1 = (pic cs:m3b) in let \p2 = (pic cs:vert) in node (d2) at (\x2,\y1) {$\af c$};
        \path let \p1 = (pic cs:m4d) in let \p2 = (pic cs:vert) in node (d3) at (\x2,\y1) {$\ag(\neg \ag\af t)$};
        \path let \p1 = (pic cs:m6b) in let \p2 = (pic cs:vert) in node (d4) at (\x2,\y1) {$\ag \neg {dead}$};
      \end{scope}
    
      \begin{scope}[every edge/.style={->,draw}]
        \draw[cm1] ($(mm1)+(1.7em,0.4em)$) edge[bend left,looseness=0.1] (d1.west);
        \draw[cm2] ($(mm2)+(1.7em,0.4em)$) edge[bend left,looseness=0.1] (d1.west);
        \draw[cm3] ($(mm3)+(1.7em,0.4em)$) edge[bend left,looseness=0.8] (d1.west);
        \draw[cm4] ($(mm4)+(1.7em,0.4em)$) edge[bend right, looseness=0.1] (d2.west);
        \draw[cm5] ($(mm5)+(1.7em,0.4em)$) edge[bend left, looseness=0.3] (d3.west);
        \draw[cm6] ($(mm6)+(1.7em,0.4em)$) edge[bend left, looseness=0.1] (d4.west);
      \end{scope}}
!bool turn, flag[2];
// t = "turn"; m = "ncrit>1"; c = "ncrit>0"
byte ncrit;
active [2] proctype user()                                    !\tikzmark{vert}!
{
  assert(_pid == 0 || _pid == 1);
again:
  !\tikzmark{m1a}!flag[_pid] = 1;!\tikzmark{m1b}!
  !\tikzmark{m2a}!turn = _pid;!\tikzmark{m2b}!
  !\tikzmark{m3a}!(flag[1 - _pid] == 0 || turn == 1 - _pid);!\tikzmark{m3b}!

  !\tikzmark{m4a}!ncrit++; !\tikzmark{m4b}!
  assert(ncrit == 1);  /* critical section */
  !\tikzmark{m4c}!ncrit--; !\tikzmark{m4d}!

  !\tikzmark{m5a}!flag[_pid] = 0;!\tikzmark{m5b}!
  !\tikzmark{m6a}!goto again!\tikzmark{m6b}!
}
    \end{lstlisting}
    \caption{Peterson's mutual exclusion protocol in PROMELA and learnt formulas for each deleted instruction.}
    \label{fig:peterson}
  \end{figure}

  \subsection{Quantitative evaluation}
  
  We quantitatively assess the performance of the various optimizations and {\ctl} fragments discussed previously. To do so, we further enrich the benchmark series through the use of random mutations of hard-coded KSs: these mutations may alter some states, re-route some edges, and spawn new states, as detailed in Appendix \ref{app:mutations}. We consider a total of 234 test samples, ranging from size 11 to 698 after minimization. We perform the benchmarks on a GNU/Linux Debian machine (\texttt{bullseye}) with 24 cores (Intel(R) Xeon(R) CPU E5-2620 @ 2.00GHz) and 256Go of RAM, using version \texttt{4.8.10} of \texttt{libz3} and \texttt{1.0} of \texttt{LearnCTL}.
  
  Table \ref{fig:table_bench} displays a summary of these benchmarks: $\beta$ stands for the refined approximation of the recurrence diameter described in Section \ref{sec:approx_acyclic}; $\neg$, for the embedding of negations in the syntactic tree introduced in Section \ref{sec:embed_negation}. The average size of the syntactic DAGs learnt is $4.14$.
  
  Option $\beta$ yields the greatest improvement, being on average at least 6 times faster than the default configuration; option $\neg$ further divides the average runtime by at least 2. These two optimizations alone speed up the average runtime by a factor of 12 to 20. The {\ctl} fragment used, all other options being equal, does not influence the average runtime as much (less than twofold in the worst case scenario); $(\ctlu, \beta, \neg)$ is the fastest option, closely followed by $(\ctla, \beta, \neg)$.
  
  Intuitively, approximating the recurrence diameter aggressively cuts down the number of {\sat} variables needed: assuming that $\alpha$ has upper bound $d$, we only need $n \cdot |Q| \cdot d$ Boolean variables $(\rank_{i,q}^u)$ instead of $n \cdot |Q|^2$. Moreover, embedding negations, despite requiring more complex clauses, results in smaller syntactic DAGs with ``free'' negations, hence faster computations, keeping in mind that the last {\sat} instances are the most expensive to solve, being the largest.
  
  \smallskip
  {\centering
  \begin{minipage}{0.47\linewidth}
    \begin{figure}[H]
      \centering
      \setlength{\tabcolsep}{5pt}
      \renewcommand{\arraystretch}{1.25}
      \scalebox{0.9}{\begin{tabular}{|c|c|c|c|} 
        \hline
        & - & $\beta$ & $\beta, \neg$ \\
        \hline
        $\ctla$ & 50 | 46870 & 14 | 6493 & 4 | 2271 \\ 
        $\ctl$ & 50 | 42658 & 8 | 5357 & 5 | 3370 \\
        $\ctlu$ & 46 | 31975 & 28 | 5064 & 4 | 1987 \\
        \hline
      \end{tabular}} 
      \caption{Number of timeouts at ten minutes | arithmetic mean (in milliseconds) on the 178 samples that never timed out of various options and fragments.}
      \label{fig:table_bench}
    \end{figure}
  \end{minipage}
  \hfill
  \begin{minipage}{0.49\linewidth}
    Figure \ref{fig:logplot_bench} further displays a log-log plot comparing the runtime of the most relevant fragments and options to $(\ctlu, \beta, \neg)$. For a given set of parameters, each point stands for one of the 234 test samples. Points above the black diagonal favour $(\ctlu, \beta, \neg)$; points below, the aforementioned option. Points on the second dotted lines at the edges of the figure represent timeouts.
  \end{minipage}}
  \medskip

  Unsurprisingly, $(\ctla, \beta, \neg)$ and $(\ctl, \beta, \neg)$ outperform $(\ctlu, \beta, \neg)$ when a minimal distinguishing formula using the operator $\au$ exists: the duality relation between $\au$ and $\eu$ is complex and, unlike the other operators, cannot be handled at no cost by the embedded negation as it depends on the \emph{release} operator.
  
  \begin{figure}
    \scalebox{0.5}{\input{best_vs_rest.tex}} 
    \caption{Comparing $(\ctlu, \beta, \neg)$ to other options on every sample.}
    \label{fig:logplot_bench}
  \end{figure}

  \section{Conclusion and Further Developments}
  
  We explored in this article the {\ctl} learning problem: we first provided a direct explicit construction before relying on a {\sat} encoding inspired by bounded model-checking to iteratively find a minimal answer. We also introduced in Section \ref{sec:learning} an explicit answer to the learning problem that belongs to the fragment $\ctl(\neg, \land, \lor, \ax, \ex)$. It remains to be seen if a smaller formula can be computed using a more exhaustive selection of $\ctl$ operators. A finer grained explicit solution could allow one to experiment with a top-down approach as well.
  
  Moreover, we provided a dedicated \texttt{C++} implementation, and evaluated it on models of higher-level formalisms such as PROMELA. Since the resulting KSs have large state spaces, further symbolic approaches are to be considered for future work, when dealing with program graphs instead of Kripke structures. In this setting, one might also consider the synthesis problem of the relevant atomic propositions from the exposed program variables. Nevertheless, the experiments on Kripke structures already showcase the benefits of the approximated recurrence diameter computation and of our extension of the syntactic DAG definition, as well as the limited relevance of the target {\ctl} fragment.
  
  Another avenue for optimizations can be inferred from the existing {\sat}-based {\ltl} learning literature: in particular, Rienier et al.~\cite{Riener19} relied on a topology-guided approach by explicitly enumerating the possible shapes of the syntactic DAG and solving the associated {\sat} instances in parallel. Given the small size on average of the formulas learnt so far and the quadratic factor impacting the number of semantic clauses such as \ref{sat:system:all_until} due to the structural variables $\leftv_{i, j}$ and $\rightv_{i, k}$, this approach could yield huge performance gains in {\ctl}'s case as well.
  
  We relied on \texttt{Z3}'s convenient \texttt{C++} API, but intuit that we would achieve better performance with state-of-the-art SAT solvers such as the winners of the yearly SAT competition~\cite{SatComp23}. We plan on converting our Boolean encoding to the \texttt{DIMACS} CNF format in order to interface our tool with modern SAT solvers.
  
  Finally, it is known that the bounded learning problem is {\np}-complete, but we would also like to find the exact complexity class of the minimal {\ctl} learning problem. We intuit that it is not, Kripke structures being a denser encoding in terms of information than lists of linear traces: as an example, one can trivially compute a {\ltl} formula (resp. a {\ctl} formula) of polynomial size that distinguishes a sample of ultimately periodic words (resp. of finite computation trees with lasso-shaped leaves), but the same cannot be said of a sample of Kripke structures. It remains to be seen if this intuition can be confirmed or infirmed by a formal proof.

  \bibliographystyle{splncs04}
  \bibliography{references}

  \newpage
  
  \appendix
  
  \section{Proof of Theorem \ref{th:bounded_sem}}
  \label{app:proof_bmc}
  
  \paragraph{$\af$.}
  Assume that $q \models \af \varphi$. Let us prove that $q \models \af^{\rdiameter(q)} \varphi$. Consider a run $r = (s_i) \in \runs(q)$. By hypothesis, we can define the index $j = \min \{i \in \integers \mid s_i \models \varphi\}$.
  
  Now, assume that $j > \rdiameter(q)$. By definition of the recurrence diameter $\alpha$, $\exists k_1, k_2 \in [0 \isep j-1]$ such that $k_1 \leq k_2$ and $s_{k_1} = s_{k_2}$. Consider the finite runs $u = (s_i)_{i \in [0 \isep k_1]}$ and $v = (s_i)_{i \in [k_1+1 \isep k_2]}$. We define the infinite, ultimately periodic run $r' = u \cdot v^\omega = (s'_i)$. By definition of $j$, $\forall i \in \integers$, $s'_i \not\models \varphi$ in order to preserve the minimality of $j$. Yet $r' \in \runs(q)$ and $q \models \af \varphi$. By contradiction, $j \leq \rdiameter(q)$. As consequence, $(q \models \af \varphi) \implies (q \models \af^{\rdiameter(q)} \varphi)$ holds.
  
  Trivially, $(q \models \af^{\rdiameter(q)} \varphi) \implies (q \models \af \varphi)$ holds. Hence, we have proven both sides of the desired equivalence for $\af$.
  
  \paragraph{$\ag$.}
  Assume that $q \models \ag^{\rdiameter(q)} \varphi$. Let us prove that $q \models \ag \varphi$. Consider a run $r = (s_i) \in \runs(q)$ and $j \in \integers$. Let us prove that $s_j \models \varphi$.
  
  State $s_j$ is obviously reachable from $q$. Let us consider a finite run without repetition $u = (s'_i)_{i \in [0 \isep k]}$ such that $s_0 = q$ and $s'_k = s_j$. By definition of the recurrence diameter, $k \leq \alpha(q)$. We define the infinite runs $v = (s_i)_{i > j}$ and $r' = u \cdot v$. Since $r' \in \runs(q)$ and $q \models \ag^{\rdiameter(q)} \varphi$, $s_k \models \varphi$, hence $s_j \models \varphi$. As a consequence, $(q \models \ag^{\rdiameter(q)} \varphi) \implies (q \models \ag \varphi)$.
  
  Trivially, $(q \models \ag \varphi) \implies (q \models \ag^{\rdiameter(q)} \varphi)$ holds. Hence, we have proven both sides of the desired equivalence for $\ag$.
  
  \paragraph{$\ef$ and $\eg$.} Formula $\ef \varphi$ (rep. $\eg \varphi$) being equivalent to the dual formula $\neg \ag \neg \varphi$ (resp. $\neg \af \neg \varphi$), the previous proofs immediately yield the desired equivalences.
  
  \paragraph{$\au$ and $\eu$.} We can handle the case of $\forall \varphi \until \psi$ in a manner similar to $\af$: we prove by contradiction that the first occurrence of $\psi$ always happens in less than $\rdiameter(q)$ steps. And the semantic equivalence for $\exists \varphi \until \psi$ can be handled in a fashion similar to $\ag$: an existing infinite run yields a conforming finite prefix without repetition of length lesser than or equal to $\rdiameter(q)$.

  \section{Proof of Theorem \ref{th:explicit_sol}}
  \label{app:proof_explicit_sol}
    
  Given two dissimilar states $q_1, q_2 \in Q$, let us prove by induction on the characteristic number $c_{q_1, q_2} = \charnum_\kripke(\{q_1\}, \{q_2\})$ that $q_1 \models D_{q_1, q_2}$ and $q_2 \not\models D_{q_1, q_2}$.
  \begin{description}
    \item[Base case.] If $c_{q_1, q_2} = 0$, then by definition $\lambda(q_1) \neq \lambda(q_2)$ and by design $D_{q_1, q_2}$ is a literal (i.e. an atomic proposition or the negation thereof) verified by $q_1$ but not by $q_2$.
    
    \item[Inductive case.] Assume that the property holds for all characteristic numbers smaller than or equal to $c \in \integers$. Consider two states $q_1, q_2 \in Q$ such that $c_{q_1, q_2} = c+1$. By definition of the refined equivalence relation $\sim_{c+1}$, $\exists q_1' \in \delta(q_1)$, $\forall q_2' \in \delta(q_2)$, $q_1' \not\sim_c q_2'$, hence $c_{q'_1, q'_2} \leq c$.
    
    By induction hypothesis, $D_{q'_1, q'_2}$ is well-defined, $q'_1 \models D_{q'_1, q'_2}$ and $q'_2 \not\models D_{q'_1, q'_2}$. As a consequence, $D_{q_1, q_2}$ is well-defined, $q_1 \models \ex \, \biggl( \underset{q_2' \in \delta(q_2)}{\bigwedge} D_{q_1', q'_2} \biggr)$, and $q_2 \not\models \ex \, \biggl( \underset{q_2' \in \delta(q_2)}{\bigwedge} D_{q_1', q'_2} \biggr)$.
    
    We handle the case where $\exists q_2' \in \delta(q_2)$, $\forall q_1' \in \delta(q_1)$, $q_1' \not\sim_c q_2'$ in a similar fashion. As a consequence, the property holds at rank $c+1$.
  \end{description}
  
  Therefore, for each $q^+ \in \psample$ and each $q^- \in \nsample$, $q^+ \models D_{q^+, q^-}$ and $q^- \not\models D_{q^+, q^-}$. Hence, $q^+ \models \sepfor_\kripke(\sample)$ and $q^- \not\models \sepfor_\kripke(\sample)$. \hfill\qed

  \section{Proof of Corollary \ref{cor:size_explicit_sol}}
  \label{app:proof_size_explicit}
  
  First, given $q^+ \in \psample$ and $q^- \in \nsample$, let us bound the size of $D_{q^+, q^-}$ based on their characteristic number $c_{q_1, q_2} = \charnum_\kripke(\{q_1\}, \{q_2\})$.
  \begin{align*}
    c_{q_1, q_2} = 0 \implies & |D_{q^+, q^-}| \leq 2 & \text{as } \lambda(q_1) \neq \lambda (q_2) \\
    c_{q_1, q_2} \geq 1 \implies & |D_{q^+, q^-}| \leq (k+1) +
    \underset{q'_2 \in \delta(q_2)}{\sum} |D_{q'_1, q'_2}| & \text{for some } q'_1 \in \delta(q_1) \\
    & \text{or } |D_{q^+, q^-}| \leq (k+1) +
    \underset{q'_1 \in \delta(q_1)}{\sum} |D_{q'_1, q'_2}| & \text{for some } q'_2 \in \delta(q_2)
  \end{align*}
  
  We are looking for an upper bound $(U_n)_{n \geq 0}$ such that $\forall n \in \integers$, $\forall q^+ \in \psample$, $\forall q^- \in \nsample$, if $c_{q^+, q^-} \leq n$, then $|D_{q^+, q^-}| \leq U_n$. We define it inductively:
  \begin{align*}
    U_0 & = 2 \\
    U_{n+1} & = k \cdot U_n + k + 1
  \end{align*}
  
  Assuming $k \geq 2$, we explicit the bound $U_n = (2 + \frac{k+1}{k-1}) \cdot k^n - \frac{k+1}{k-1} \leq 5 \cdot k^n$. As $(\{q^+\}, \{q^-\})$ is a sub-sample of $S$, $c_{q^+, q^-}\leq c$ and $|D_{q^+, q^-}| \leq U_c \leq 5 \cdot k^c$. We can finally bound the size of $\sepfor_\kripke(\sample)$:
  \begin{align*}
    |\sepfor_\kripke(\sample)| & \leq (|\psample|-1) \cdot (|\nsample|-1) +
    \underset{\substack{q^+ \in \psample \\ q^- \in \nsample}}{\sum} |D_{q^+, q^-}| \\
    & \leq |\psample| \cdot |\nsample| + |\psample| \cdot |\nsample| \cdot U_c \\
    & \leq (5 \cdot k^c + 1) \cdot |\psample| \cdot |\nsample|
  \end{align*}
  
  Yielding the aforementioned upper bound. If $k=1$, then $U_n = 2 \cdot n + 2$ and the rest of the proof is similar to the previous case. \hfill\qed

  \section{Encoding the Syntactic DAG}
  \label{app:sat_tree}
  
  The clauses encoding the syntactic DAG are the following:
  
  {\allowdisplaybreaks
  \small
  \begin{align}
    \underset{i \in [1 \isep n]}{\bigwedge} \Biggl[
    \Biggl(\underset{o \in \labels}{\bigvee} \tau_i^o\Biggr)
    \land \Biggl(\underset{\substack{o_1, o_2 \in \labels \\ o_1 \neq o_2}}{\bigwedge} [\neg \tau_i^{o_1} \lor \neg \tau_i^{o_2}]\Biggr)
    \Biggr]
    \tag{$\mathtt{xor}_\tau$}
    \label{sat:tree:unique_label} \\
    \underset{i \in [1 \isep n]}{\bigwedge} \Biggl[
    \Biggl(\underset{j \in [0 \isep i-1]}{\bigvee} \leftv_{i, j}\Biggr)
    \land \Biggl(\underset{\substack{j, k \in [0 \isep i-1] \\ j < k}}{\bigwedge} [\neg \leftv_{i, j} \lor \neg \leftv_{i, k}]\Biggr)
    \Biggr]
    \tag{$\mathtt{xor}_l$}
    \label{sat:tree:unique_left} \\
    \underset{i \in [1 \isep n]}{\bigwedge} \Biggl[
    \Biggl(\underset{j \in [0 \isep i-1]}{\bigvee} \rightv_{i, j}\Biggr)
    \land \Biggl(\underset{\substack{j, k \in [0 \isep i-1] \\ j < k}}{\bigwedge} [\neg \rightv_{i, j} \lor \neg \rightv_{i, k}]\Biggr)
    \Biggr]
    \tag{$\mathtt{xor}_r$}
    \label{sat:tree:unique_right} \\
    \underset{\substack{i \in [1 \isep n] \\ a \in \ap \cup \{\top\}}}{\bigwedge}
    [\tau_i^{a} \implies (\leftv_{i, 0} \land \rightv_{i, 0})]
    \tag{$\mathtt{ar}_0$}
    \label{sat:tree:leaf_no_children} \\
    \underset{\substack{i \in [1 \isep n] \\ o \in \{\neg, \ax, \af, \ag\}}}{\bigwedge}
    [\tau_i^{o} \implies (\leftv_{i, 0} \land \neg \rightv_{i, 0})]
    \tag{$\mathtt{ar}_1$}
    \label{sat:tree:unary_one_child} \\
    \underset{\substack{i \in [1 \isep n] \\ o \in \{\land, \lor, \au\}}}{\bigwedge}
    [\tau_i^{o} \implies (\neg \leftv_{i, 0} \land \neg \rightv_{i, 0})]
    \tag{$\mathtt{ar}_2$}
    \label{sat:tree:binary_two_children}
  \end{align}}
  
  Note in particular that clauses \ref{sat:tree:unique_label} to \ref{sat:tree:binary_two_children} do not encode a single syntactic DAG of size $n$ as they do not guarantee connectivity, but instead a forest of DAGs, as shown in Figure \ref{fig:forest_dags}. However, we will only keep the DAG rooted in node $n$, being the only one later clause \ref{sat:system:sample} enforces semantic constraints on. Figure \ref{fig:forest_dags} displays a forest of DAGs resulting in formula $\varphi = \forall a \until \neg b$ rooted in node $7$.
  
  \begin{figure}[h]
    \centering
    \begin{tikzpicture}[x=1.5cm, y=0.75cm]
      \node (aw) at (0.5,2) {$\au_{[7]}$};
      \node (a) at (0,1) {$a_{[5]}$};
      \node (neg) at (1,1) {$\neg_{[3]}$};
      \node (b) at (1,0) {$b_{[2]}$};
      \draw[->] (aw) -- (a);
      \draw[->] (aw) -- (neg);
      \draw[->] (neg) -- (b);
      
      \node (ex) at (2,1) {$\ax_{[4]}$};
      \node (c) at (2,0) {$c_{[1]}$};
      \draw[->] (ex) -- (c);
      
      \node (top) at (3,0) {$\top_{[6]}$};
    \end{tikzpicture}
    \caption{A valid forest of syntactic DAGs of size $n = 7$. The exponent stands for the index of each node.}
    \label{fig:forest_dags}
  \end{figure}

  \section{Encoding the Boolean Operators}
  \label{app:sat_bool}
  
  The clauses encoding the semantics of the Boolean operators are the following:
  {\allowdisplaybreaks
  \small
  \begin{align}
    \underset{\substack{i \in [2 \isep n] \\ k \in [1 \isep i-1]}}{\bigwedge} & \Biggl[
    (\tau_i^\neg \land \rightv_{i, k}) \implies
    \underset{q \in Q}{\bigwedge} (\varphi_i^{q} \iff \neg \varphi_k^{q})
    \Biggr]
    \tag{$\mathtt{sem}_{\neg}$}
    \label{sat:system:not} \\
    \underset{\substack{i \in [2 \isep n] \\ j, k \in [1 \isep i-1]}}{\bigwedge} & \Biggl[
    (\tau_i^\land \land \leftv_{i, j} \land \rightv_{i, k}) \implies
    \underset{q \in Q}{\bigwedge} (\varphi_i^{q} \iff [\varphi_j^q \land \varphi_k^q])
    \Biggr]
    \tag{$\mathtt{sem}_{\land}$}
    \label{sat:system:and} \\
    \underset{\substack{i \in [2 \isep n] \\ j, k \in [1 \isep i-1]}}{\bigwedge} & \Biggl[
    (\tau_i^\lor \land \leftv_{i, j} \land \rightv_{i, k}) \implies
    \underset{q \in Q}{\bigwedge} (\varphi_i^{q} \iff [\varphi_j^q \lor \varphi_k^q])
    \Biggr]
    \tag{$\mathtt{sem}_{\lor}$}
    \label{sat:system:or}
  \end{align}}

  \section{A Spurious Fixed Point}
  \label{app:wrong_semantics}

  The semantics of the temporal operators $\af$, $\ag$ and $\au$ are significantly harder to model than $\ax$ or the Boolean connectors: consider that the inductive property $(q \models \ag a) \iff (q \models a) \land \underset{q' \in \delta(q)}{\bigwedge} (q' \models \ag a)$ holds. Yet the naive encoding $(\tau_2^{\ag} \land \rightv_{2, 1}) \implies \underset{q \in Q}{\bigwedge} \Biggl(\varphi_2^q \iff \biggl[\varphi_1^q \land \underset{q' \in \delta(q)}{\bigwedge} \varphi_2^{q'}\biggr]\biggr)$ may in some cases result in a spurious fixed point, as depicted in Figure \ref{fig:false_sem}. 
  
  \begin{figure}
    \centering
    \begin{tikzpicture}[x=1cm, y=0.5cm]
      \node[minimum size=0pt] (k) at (-1.5,0) {$\kripke$};
      \node[state, label={180:$\mathbf{q}_1$}, minimum size=0pt] (q1) at (0,0) {$a$};
      \node[state, label={0:$\mathbf{q}_2$}, minimum size=0pt] (q2) at (1,1) {$a$};
      \node[state, label={0:$\mathbf{q}_3$}, minimum size=0pt] (q3) at (1,-1) {$a$};
      \draw[->] (q1) -- (q2);
      \draw[->] (q2) -- (q3);
      \draw[->] (q3) -- (q1);
      
      \node (ag) at (3,1) {$\ag_{[2]}$};
      \node (a) at (3,-1) {$a_{[1]}$};
      \draw[->] (ag) -- (a);
      
      \node (phi1) at (5,0) [align=left]{$\varphi_1^{q_1} = 1$ \\ $\varphi_1^{q_2} = 1$ \\ $\varphi_1^{q_3} = 1$};
      \node (phi2) at (6.5,0) [align=left]{$\varphi_2^{q_1} = 0$ \\ $\varphi_2^{q_2} = 0$ \\ $\varphi_2^{q_3} = 0$};
    \end{tikzpicture}
    \caption{Naive semantics result in the incorrect claim that $q_1 \not\models \ag a$.}
    \label{fig:false_sem}
  \end{figure}
  
  Intuitively, while every state of $\kripke$ verifies the atomic proposition $a$, each state $q$ that belongs to the loop can claim it does not verify $\ag a$ (i.e. $\varphi_2^q = 0$) as its successor $q'$ claims it does not either (since $\varphi_2^{q'}$ = 0), hence complying with the naive semantic equivalence clause outlined previously.

  \section{Encoding the Universal Operators}
  \label{app:sat_universal_ops}
  
  The clauses encoding the semantics of the temporal operators $\ag$ and $\af$ are the following:
  {\allowdisplaybreaks
  \small
  \begin{align}
    \underset{\substack{i \in [2 \isep n] \\ k \in [1 \isep i-1]}}{\bigwedge} & \Biggl[
    (\tau_i^{\af} \land \rightv_{i, k}) \implies \nonumber \\
    & \underset{\substack{q \in Q \\ u \in [1 \isep \rdiameter(q)]}}{\bigwedge} \Biggl(\rank_{i,q}^u \iff \Biggl[\varphi_k^q \lor \underset{q' \in \delta(q)}{\bigwedge} \rank_{i, q'}^{\min(\rdiameter(q'), u-1)}\Biggr]\Biggr)
    \Biggr]
    \tag{$\mathtt{sem}_{\af}$}
    \label{sat:system:all_finally} \\
    \underset{\substack{i \in [2 \isep n] \\ k \in [1 \isep i-1]}}{\bigwedge} & \Biggl[
    (\tau_i^{\ag} \land \rightv_{i, k}) \implies \nonumber \\
    & \underset{\substack{q \in Q \\ u \in [1 \isep \rdiameter(q)]}}{\bigwedge}
    \Biggl(\rank_{i,q}^u \iff \Biggl[\varphi_k^q \land \underset{q' \in \delta(q)}{\bigwedge} \rank_{i, q'}^{\min(\rdiameter(q'), u-1)}\Biggr]\Biggr)
    \Biggr]
    \tag{$\mathtt{sem}_{\ag}$}
    \label{sat:system:all_globally}
  \end{align}}

  \section{Encoding the Existential Operators}
  \label{app:sat_existential_ops}
  
  The existing syntactic clauses \ref{sat:tree:unique_label}, \ref{sat:tree:unary_one_child}, and \ref{sat:tree:binary_two_children} as well as the clauses \ref{sat:system:rank_semantics}, \ref{sat:system:rank_ascent}, \ref{sat:system:rank_descent}, \ref{sat:system:rank_forbidden}, and \ref{sat:system:rank_start} regulating the bounded semantics should be updated to take the existential operators $\ex$, $\ef$, $\eg$, and $\eu$ into account. Then the following clauses encoding the existential semantics should be added:
  {\allowdisplaybreaks
  \small
  \begin{align}
    \underset{\substack{i \in [2 \isep n] \\ k \in [1 \isep i-1]}}{\bigwedge} \Biggl[
    & (\tau_i^{\ex} \land \rightv_{i, k}) \implies
    \underset{q \in Q}{\bigwedge} \Biggl(\varphi_i^{q} \iff \underset{q' \in \delta(q)}{\bigvee} \varphi_k^{q'}\Biggr)
    \Biggr]
    \tag{$\mathtt{sem}_{\ex}$}
    \label{sat:system:exists_next} \\
    \underset{\substack{i \in [2 \isep n] \\ k \in [1 \isep i-1]}}{\bigwedge} & \Biggl[
    (\tau_i^{\ef} \land \rightv_{i, k}) \implies \nonumber \\
    & \underset{\substack{q \in Q \\ u \in [1 \isep \rdiameter(q)]}}{\bigwedge} \Biggl(\rank_{i,q}^u \iff \Biggl[\varphi_k^q \lor \underset{q' \in \delta(q)}{\bigvee} \rank_{i, q'}^{\min(\rdiameter(q'), u-1)}\Biggr]\Biggr)
    \Biggr]
    \tag{$\mathtt{sem}_{\ef}$}
    \label{sat:system:exists_finally} \\
    \underset{\substack{i \in [2 \isep n] \\ k \in [1 \isep i-1]}}{\bigwedge} & \Biggl[
    (\tau_i^{\eg} \land \rightv_{i, k}) \implies \nonumber \\
    & \underset{\substack{q \in Q \\ u \in [1 \isep \rdiameter(q)]}}{\bigwedge} \Biggl(\rank_{i,q}^u \iff \Biggl[\varphi_k^q \land \underset{q' \in \delta(q)}{\bigvee} \rank_{i, q'}^{\min(\rdiameter(q'), u-1)}\Biggr]\Biggr)
    \Biggr]
    \tag{$\mathtt{sem}_{\eg}$}
    \label{sat:system:exists_globally} \\
    \underset{\substack{i \in [2 \isep n] \\ j, k \in [1 \isep i-1]}}{\bigwedge} & \Biggl[
    (\tau_i^{\eu} \land \leftv_{i, j} \land \rightv_{i, k}) \implies \nonumber \\
    & \underset{\substack{q \in Q \\ u \in [1 \isep \rdiameter(q)]}}{\bigwedge} \Biggl(\rank_{i,q}^u \iff \Biggl[\varphi_k^q \lor \Biggl(\varphi_j^q \land \underset{q' \in \delta(q)}{\bigvee} \rank_{i, q'}^{\min(\rdiameter(q'), u-1)}\Biggr)\Biggr]\Biggr)
    \Biggr]
    \tag{$\mathtt{sem}_{\eu}$}
    \label{sat:system:exists_until}
  \end{align}}

  \section{Updated Clauses with Embedded Negations}
  \label{app:embedded_neg}
  
  We consider the following updated semantic clauses:
  {\allowdisplaybreaks
  \small
  \begin{align}
    \underset{\substack{i \in [2 \isep n] \\ j \in [1 \isep i-1]}}{\bigwedge} \Biggl[
    & (\tau_i^{\ax} \land \leftv_{i, j}) \implies
    \underset{q \in Q}{\bigwedge} \Biggl([(\neg \nu_i \land \neg\varphi_i^q) \lor (\nu_i \land \varphi_i^q)] \iff \underset{q' \in \delta(q)}{\bigwedge} \varphi_j^{q'}\Biggr)
    \Biggr]
    \tag{$\tilde{\mathtt{sem}}_{\ax}$}
    \label{sat:system:neg:all_next} \\
    \underset{\substack{i \in [1 \isep n] \\ q \in Q}}{\bigwedge} &
    (\tau_i^\top \implies [(\neg \nu_i \land \neg\varphi_i^q) \lor (\nu_i \land \varphi_i^q)])
    \tag{$\tilde{\mathtt{sem}}_\top$}
    \label{sat:system:neg:top} \\
    \underset{\substack{i \in [1 \isep n] \\ q \in Q}}{\bigwedge} & \Biggl[
    \biggl(\underset{a \in \lambda(q)}{\bigwedge} (\tau_i^a \implies [(\neg \nu_i \land \neg\varphi_i^q) \lor (\nu_i \land \varphi_i^q)]) \biggr) \nonumber \\
    & \land \biggl(\underset{a \not\in \lambda(q)}{\bigwedge} (\tau_i^a \implies \neg [(\neg \nu_i \land \neg\varphi_i^q) \lor (\nu_i \land \varphi_i^q)]) \biggr)
    \Biggr]
    \tag{$\tilde{\mathtt{sem}}_a$}
    \label{sat:system:neg:predicates} \\
    \underset{\substack{i \in [2 \isep n] \\ j, k \in [1 \isep i-1]}}{\bigwedge} & \Biggl[
    (\tau_i^\land \land \leftv_{i, j} \land \rightv_{i, k}) \implies
    \underset{q \in Q}{\bigwedge} ([(\neg \nu_i \land \neg\varphi_i^q) \lor (\nu_i \land \varphi_i^q)] \iff [\varphi_j^q \land \varphi_k^q])
    \Biggr]
    \tag{$\tilde{\mathtt{sem}}_\land$}
    \label{sat:system:neg:and} \\
    \underset{\substack{i \in [2 \isep n] \\ j, k \in [1 \isep i-1]}}{\bigwedge} & \Biggl[
    (\tau_i^\lor \land \leftv_{i, j} \land \rightv_{i, k}) \implies
    \underset{q \in Q}{\bigwedge} ([(\neg \nu_i \land \neg\varphi_i^q) \lor (\nu_i \land \varphi_i^q)] \iff [\varphi_j^q \lor \varphi_k^q])
    \Biggr]
    \tag{$\tilde{\mathtt{sem}}_\lor$}
    \label{sat:system:neg:or} \\
    \underset{i \in [2 \isep n]}{\bigwedge} & \Biggl[
    \Biggl(\underset{o \in \{\af, \ag, \au\}}{\bigwedge} \neg \tau_i^o\Biggr) \implies
    \underset{q \in Q}{\bigwedge} ([(\neg \nu_i \land \neg\varphi_i^q) \lor (\nu_i \land \varphi_i^q)] \iff\rank_{i,q}^{\rdiameter(q)})
    \Biggr]
    \tag{$\tilde{\mathtt{sem}}_\rho$}
    \label{sat:system:neg:rank_semantics}
  \end{align}}

  \section{Generating a test sample with mutations}
  \label{app:mutations}

  Given a KS $\kripke = (Q, \delta, \lambda)$ and $k \in \integers^*$, we design a $k$-\emph{mutant} $\kripke' = (Q', \delta', \lambda')$ of $\kripke$ by randomly applying these mutation rules $k$ times to a copy of $\kripke$, as shown in Figure \ref{fig:mutations}:
  \begin{description}
    \item[Mutating states.] Pick a random state $q \in Q$ and randomly pick a new label $\lambda'(q) \in 2^\ap$ such that $\lambda'(q) \neq \lambda(q)$.
    
    \item[Mutating outgoing edges.] Pick two random states $q \in Q$ and $q' \in \delta(q)$, remove $q'$ from $\delta'(q)$, and add instead if it exists a new destination $q''$ to $\delta'(q)$ such that $q'' \not\in \delta(q)$.
    
    \item[Creating mutant states.] Pick two random states $q \in Q$ and $q' \in \delta(q)$, remove $q'$ from $\delta'(q)$, add a new state $q''$ to $Q'$ such that $q'' \in \delta'(q)$, $q' \in \delta'(q'')$, and $\lambda'(q'')$ is a random element of $2^\ap$
  \end{description}
  
  \begin{figure}[h!]
    \centering
    \begin{tikzpicture}[x=1cm, y=0.5cm]
      \node[minimum size=0pt] (q1) at (-2,0) {$\kripke$};
      
      \node[state, label={90:$\mathbf{q}$}, minimum size=0pt] (q1) at (0,0) {$a$};
      
      \node[state, label={90:$\mathbf{q}$}, minimum size=0pt] (q2) at (2,0) {};
      \node[state, label={90:$\mathbf{q'}$}, minimum size=0pt] (q2p) at (3,1) {};
      \node[state, label={90:$\mathbf{q''}$}, minimum size=0pt] (q2pp) at (3,-1) {};
      \draw[->] (q2) -- (q2p);
      
      \node[state, label={90:$\mathbf{q}$}, minimum size=0pt] (q3) at (5,0) {};
      \node[state, label={90:$\mathbf{q'}$}, minimum size=0pt] (q3p) at (7,0) {};
      \draw[->] (q3) -- (q3p);
    \end{tikzpicture}
    
    \bigskip
    \begin{tikzpicture}[x=1cm, y=0.5cm]
      \node[minimum size=0pt] (q1) at (-2,0) {$\kripke'$};
      
      \node[state, label={90:$\mathbf{q}$}, minimum size=0pt] (q1) at (0,0) {$b$};
      
      \node[state, label={90:$\mathbf{q}$}, minimum size=0pt] (q2) at (2,0) {};
      \node[state, label={90:$\mathbf{q'}$}, minimum size=0pt] (q2p) at (3,1) {};
      \node[state, label={90:$\mathbf{q''}$}, minimum size=0pt] (q2pp) at (3,-1) {};
      \draw[->] (q2) -- (q2pp);
      
      \node[state, label={90:$\mathbf{q}$}, minimum size=0pt] (q3) at (5,0) {};
      \node[state, label={90:$\mathbf{q'}$}, minimum size=0pt] (q3p) at (7,0) {};
      \node[state, label={90:$\mathbf{q''}$}, minimum size=0pt] (q3pp) at (6,0) {};
      \draw[->] (q3) -- (q3pp);
      \draw[->] (q3pp) -- (q3p);
    \end{tikzpicture}
    
    \caption{The original KS $\kripke$ and the resulting effect of mutation rules on $\kripke'$.}
    \label{fig:mutations}
  \end{figure}

\end{document}